\documentclass{article}
\usepackage{amssymb}

\usepackage{graphicx}
\usepackage{amsmath}

\input{tcilatex}  

\begin{document}

\title{\textbf{Surveillance on the light-front gauge fixing Lagrangians}}
\author{Alfredo T.Suzuki$^{a}$ and J.H.O.Sales$^{b}$ \\
%EndAName
$^{a}$Instituto de F\'{i}sica Te\'{o}rica-UNESP, \\
01405-900 S\~{a}o Paulo, Brazil.\\
$^{b}$Faculdade de Tecnologia de S\~{a}o Paulo-DEG, \\
P\c{c}a. Cel. Fernando Prestes, \\
01124-060 S\~{a}o Paulo, SP, Brazil}
\maketitle

\begin{abstract}
In this work we propose \emph{two} Lagrange multipliers with distinct
coefficients for the light-front gauge that leads to the complete
(non-reduced) propagator. This is accomplished via $(n\cdot A)^2+(\partial
\cdot A)^2$ terms in the Lagrangian density. These lead to a well-defined
and exact though Lorentz non invariant light front propagator.
\end{abstract}

\section{Introduction}

Traditional (at the classical level) gauge fixing in the light front is done
by adding a Lagrange multiplier of the form $(n\cdot A)^{2}$, into the
Lagrangian density, where $n_{\mu }$ is the external light-like vector,
i.e., $n^{2}=0$, and $A_{\mu }$ is the vector potential (eventual color
indices omitted for brevity). This leads to the usual two term light-front
gauge propagator with the characteristic $(k\cdot n)^{-1}$ pole. However,
when canonical quantization of the theory is carried out in the light front,
it has been pointed out since long times past \cite{soper,tomboulis} as well
as more recently \cite{prem}, that there is an emergence of a third term
with a double non-local pole, prominently absent in the classical
(traditional) derivation of the propagator above referred to. This term has
always been systematically omitted on various grounds \cite{soper,tomboulis}
and the reduced two-term propagator has been since there the working
propagator for relevant computations in the light front milieu. Our present
work is to point out a solution to the apparent anomaly present when we
compare the classical derivation with the canonical quantization in the
light-front. Here we limit ourselves with this question and do not concern
ourselves in considering the relevance or not of the obtained third term -
this can more clearly be seen and shall be dealt with in particular settings
of certain physical processes in the forthcoming papers.

Our contribution in this paper is to show that the condition $n\cdot A=0$ ($%
n^{2}=0$) is \emph{necessary} but \emph{not sufficient} to define the
light-front gauge. It leads to the standard form of the light-front reduced
propagator with two terms. The \emph{necessary} and \emph{sufficient}
condition to uniquely define the light-front gauge is given by $n\cdot
A=\partial \cdot A=0$ so that the corresponding Lagrange multipliers to be
added to the Lagrangian density are proportional to $(n\cdot A)^2+(\partial
\cdot A)^2$ instead of the usual $(n\cdot A)^{2}$. Note that the condition $%
\partial \cdot A=0$ in the light-cone variables defines exactly (for $n\cdot
A=A^{+}=0$) the constraint 
\begin{equation}  \label{constraint}
A^{-}=\frac{\partial ^{\perp }A^{\perp }}{\partial ^{+}}\Rightarrow \frac{
k^{\perp }A^{\perp }}{k^{+}}.
\end{equation}
This constraint, together with $A^{+}=0$, once substituted into the
Lagrangian density yields the so-called two-component formalism in the
light-front, where one is left with only physical degrees of freedom, and
Ward-Takahashi identities and multiplicative renormalizability of pure
Yang-Mills field theory is verified \cite{gluonvertex}.

\section{Massless vector field propagator}

In our previous work \cite{reex1}, we showed that a single Lagrangian
multiplier of the form $(n\cdot A)(\partial \cdot A)$ with $n\cdot
A=\partial \cdot A=0$ leads to a propagator in the light-front gauge that
has no residual gauge freedom left. However, it is clear that the constraint 
$(n\cdot A)(\partial \cdot A)=0$ does not uniquely lead to the \emph{\
necessary} conditions $n\cdot A=\partial \cdot A=0$, since the constraint is
satisfied even if only one of the factors vanish. In this sequel work we
propose a more general form with two multipliers each with its corresponding
condition so that they are uniquely defined, and show that we arrive at the
same propagator with no residual gauge freedom left.

The Lagrangian density for the vector gauge field (for simplicity we
consider an Abelian case) is given by 
\begin{equation}
\mathcal{L}=-\frac{1}{4}F_{\mu \nu }F^{\mu \nu }-\frac{1}{2\alpha }\left(
n_{\mu }A^{\mu }\right) ^{2}-\frac{1}{2\beta }\left( \partial _{\mu }A^{\mu
}\right) ^{2},  \label{lag}
\end{equation}
where $\alpha $ and $\beta $ are arbitrary constants. Of course, with these
additional gauge breaking terms, the Lagrangian density is no longer gauge
invariant and as such gauge fixing problem in this sense do not exist
anymore. Now, $\partial _{\mu }A^{\mu }$ doesn't need to be zero so that the
Lorenz condition is verified \cite{Gupta}.

The classical procedure to obtain the field propagator is to look for the
inverse operator present in the quadratic term in the Lagrangian density,
namely the one corresponding to the differential operator sandwiched between
the vector potentials. For the Abelian gauge field Lagrangian density we
have: 
\begin{equation}
\mathcal{L}=-\frac{1}{4}F_{\mu \nu }F^{\mu \nu }-\frac{1}{2\beta }\left(
\partial _{\mu }A^{\mu }\right) ^{2}-\frac{1}{2\alpha }\left( n_{\mu }A^{\mu
}\right) ^{2}=\mathcal{L}_{\text{E}}+\mathcal{L}_{GF}  \label{2}
\end{equation}

By partial integration and considering that terms which bear a total
derivative don't contribute and that surface terms vanish since $%
\lim\limits_{x\rightarrow \infty }A^{\mu }(x)=0$, we have 
\begin{equation}
\mathcal{L}_{\text{E}}=\frac{1}{2}A^{\mu }\left( \square g_{\mu \nu
}-\partial _{\mu }\partial _{\nu }\right) A^{\nu }  \label{3}
\end{equation}
and 
\begin{eqnarray}
\mathcal{L}_{GF}&=&-\frac{1}{2\beta }\partial _{\mu }A^{\mu }\partial _{\nu
}A^{\nu }-\frac{1}{2\alpha }n_{\mu }A^{\mu }n_{\nu }A^{\nu }  \notag \\
&=&\frac{1}{2\beta}A^{\mu }\partial _{\mu }\partial _{\nu }A^{\nu }-\frac{1}{
2\alpha }A^{\mu }n_{\mu }n_{\nu }A^{\nu }  \label{4}
\end{eqnarray}
so that 
\begin{equation}
\mathcal{L}=\frac{1}{2}A^{\mu }\left( \square g_{\mu \nu }-\partial _{\mu
}\partial _{\nu }+\frac{1}{\beta }\partial _{\mu }\partial _{\nu }-\frac{1}{
\alpha }n_{\mu }n_{\nu }\right) A^{\nu }  \label{5}
\end{equation}

To find the gauge field propagator we need to find the inverse of the
operator between parenthesis in (\ref{5}). That differential operator in
momentum space is given by: 
\begin{equation}
O_{\mu \nu }=-k^{2}g_{\mu \nu }+k_{\mu }k_{\nu }-\theta k_{\mu }k_{\nu
}-\lambda n_{\mu }n_{\nu }\,,  \label{6}
\end{equation}
where $\theta =\beta ^{-1}$ and $\lambda =\alpha ^{-1}$, so that the
propagator of the field, which we call $G^{\mu \nu }(k)$, must satisfy the
following equation: 
\begin{equation}
O_{\mu \nu }G^{\nu \lambda }\left( k\right) =\delta _{\mu }^{\lambda }
\label{7}
\end{equation}

$G^{\nu \lambda }(k)$ can now be constructed from the most general tensor
structure that can be defined, i.e., all the possible linear combinations of
the tensor elements that composes it \cite{progress}: 
\begin{eqnarray}
G^{\mu \nu }(k) &=&g^{\mu \nu }A+k^{\mu }k^{\nu }B+k^{\mu }n^{\nu }C+n^{\mu
}k^{\nu }D+k^{\mu }m^{\nu }E+  \notag \\
&&+m^{\mu }k^{\nu }F+n^{\mu }n^{\nu }G+m^{\mu }m^{\nu }H+n^{\mu
}m^{\nu}I+m^{\mu }n^{\nu }J  \label{8}
\end{eqnarray}
where $m^\mu$ is the light-like vector dual to the $n^\mu$, and $A$, $B$, $C$
, $D$, $E$, $F$, $G$, $H$, $I$ and $J$ are coefficients that must be
determined in such a way as to satisfy (\ref{7}). Of course, it is
immediately clear that since (\ref{5}) does not contain any external
light-like vector $m_{\mu }$, the coefficients $E=F=H=I=J=0$ straightaway.
Then, we have 
\begin{equation}
A=-(k^{2})^{-1}  \label{9}
\end{equation}
\begin{equation}
(k\cdot n)(1-\theta )G-\theta k^{2}D=0  \label{10}
\end{equation}
\begin{equation}
(-k-\lambda n^{2})G-\lambda (k\cdot n)D-\lambda A=0  \label{11}
\end{equation}
\begin{equation}
-(k^{2}+\lambda n^{2})C-\lambda (k\cdot n)B=0  \label{12}
\end{equation}
\begin{equation}
(1-\theta )A-\theta k^{2}B+(1-\theta )(k\cdot n)C=0  \label{13}
\end{equation}

From (\ref{10}) we have 
\begin{equation}
G=\frac{k^{2}}{(k\cdot n)(\beta -1)}D  \label{14a}
\end{equation}
which inserted into (\ref{11}) yields 
\begin{equation}
D=\frac{-(k\cdot n)(\beta -1)}{(\alpha k^{2}+n^{2})k^{2}+(k\cdot
n)^{2}(\beta -1)}A  \label{15}
\end{equation}

From (\ref{12}) and (\ref{13}) we obtain 
\begin{equation}
B=\frac{-(\alpha k^{2}+n^{2})}{k\cdot n}C  \label{16}
\end{equation}
and 
\begin{equation*}
C=\frac{-(\beta -1)(k\cdot n)}{(\alpha k^{2}+n^{2})k^{2}+(k\cdot
n)^{2}(\beta -1)}A=D
\end{equation*}

We have then, 
\begin{eqnarray}
G^{\mu\nu}(k)&=&-\frac{1}{k^2}\left\{g^{\mu\nu}+\frac{(\alpha
k^2+n^2)(\beta-1)}{(\alpha k^2+n^2)k^2+(k\cdot n)^2(\beta-1)}k^\mu\,k^\nu
\right.  \notag \\
&&-\frac{(\beta-1)(k^\mu n^\nu+n^\mu k^\nu)}{(\alpha k^2+n^2)k^2+(k\cdot
n)^2(\beta-1)}(k\cdot n)  \notag \\
&&-\left. \frac{1}{(\alpha k^2+n^2)k^2+(k\cdot n)^2(\beta-1)}k^2\,n^\mu
n^\nu \right\}
\end{eqnarray}

In the light-font $n^{2}=0$ and taking the limit $\alpha $,$\beta
\rightarrow 0$, we have 
\begin{equation*}
A=\frac{-1}{k^{2}}
\end{equation*}
\begin{equation*}
B=0
\end{equation*}
\begin{equation*}
C=D=\frac{1}{k^{2}(k\cdot n)}
\end{equation*}
\begin{equation*}
G=\frac{-1}{(k\cdot n)^{2}}
\end{equation*}

Therefore, the complete propagator in the light-front gauge is: 
\begin{equation}
G^{\mu \nu }(k)=-\frac{1}{k^{2}}\left\{ g^{\mu \nu }-\frac{k^{\mu }n^{\nu
}+n^{\mu }k^{\nu }}{k\cdot n}+\frac{n^{\mu }n^{\nu }}{(k\cdot n)^{2}}
k^{2}\right\} \,,  \label{17}
\end{equation}
which has the prominent third term commonly referred to as \emph{contact term%
}, oftentimes dropped on various grounds in actual calculations in the
light-front. This result of ours concides exactly with the one in \cite{prem}%
, where the presence of this term seemingly does not significantly affect
the beta function for the Yang-Mills theory and renormalization constants
satisfy the Ward-Takahashi identity $Z_1=Z_3$. Yet in other contexts this
term may prove to be crucial in the light-front formulation of the theory 
\cite{jh}.

\section{Conclusions}

We have constructed Lagrange multipliers in the light-front that leads to a
well-defined fixed gauge choice so that no unphysical degrees of freedom are
left. In other words, no residual gauge remains to be dealt with. Moreover
this allows us to get the complete (non-reduced) propagator including the
so-called contact term, so that there is no anomaly in going through to the
quantum case.

We emphasize that in \cite{reex1} the Lagrange multiplier term of the form $%
(n\cdot A)(\partial \cdot A)=0$ was such that $n\cdot A=0$ \emph{and} $%
(\partial \cdot A)=0$ simultaneously. This means that, of course, $(n\cdot
A)+(\partial \cdot A)=0$. Therefore $[(n\cdot A)+(\partial \cdot A)]^2=0$,
or 
\begin{eqnarray}
(n\cdot A)^2+(\partial \cdot A)^2+2(n\cdot A)(\partial \cdot A)&=&0  \notag
\\
(n\cdot A)^2+(\partial \cdot A)^2&=&-2(n\cdot A)(\partial \cdot A)\,,
\end{eqnarray}
thus establishing the complete equivalence between the two cases. Note that
this equivalence guarantees that we still have decoupling of the ghost
fields from the physical fields. \vspace{.5cm}

\noindent \textsc{acknowledgements}: A.T.Suzuki wishes to thank CNPq for
partial support under process 303848/2002-2 and J.H.O.Sales is supported by
Fapesp (00/09018-0).

\section{Appendix}

In this Appendix we review basic concepts of gauge invariance, gauge fixing
and gauge choice that are commonly forgotten or taken for granted, but we
deem appropriate to clarify the issues presented in this work. It is clear
that Maxwell's equations 
\begin{equation}  \label{Maxwell}
\partial_\mu F^{\mu\nu}=\partial_\mu\left(\partial^\mu A^\nu-\partial^\nu
A^\mu\right)=0,
\end{equation}
do not completely specify the vector potential $A^\mu(x)$. For, if $A^\mu(x)$
satisfies (\ref{Maxwell}), so does 
\begin{equation}  \label{gauge}
A^{^{\prime}\mu}(x)=A^\mu(x)+\partial ^\mu \Lambda(x),
\end{equation}
for any arbitrary function $\Lambda(x)$. It is also clear that both vector
potentials $A^\mu$ and $A^{^{\prime}\mu}$ yield the same electric and
magnetic fields $\vec{E}(x)$ and $\vec{B}(x)$, which are invariant under the
substitutions 
\begin{eqnarray}
A_0&\rightarrow&A^{^{\prime}}_0=A_0+\partial_0 \Lambda  \notag \\
\vec{A}&\rightarrow&\vec{A}^{\:^{\prime}}=\vec{A}-\vec{\nabla}\Lambda.
\end{eqnarray}
This lack of uniqueness of the vector potential for given electric and
magnetic fields generates difficulties when, for example, we have to perform
functional integrals over the different field configurations. This lack of
uniqueness may be reduced by imposing a further condition on $A^\mu(x)$,
besides those required by Maxwell's equations (\ref{Maxwell}). It is
customary to impose the so-called ``Lorenz condition''\footnote{%
This is not a misprint. J.D.Jackson \cite{Lorenz} calls our attention to
this giving first credit to whom it is deserved.} 
\begin{equation}  \label{Lorenz}
\partial_\mu A^\mu(x)=0,
\end{equation}
which is clearly the unique covariant condition that is linear in $A^\mu$.
However, even the imposition of the Lorenz condition does not fix the gauge
potential, since if $A$ and $A^{\prime}$ are related as in (\ref{gauge}),
then both of them will satisfy (\ref{Lorenz}) if 
\begin{equation}
\square \Lambda\equiv \partial_\mu \Lambda^\mu=0.
\end{equation}

When we choose a particular $A^{^{\prime}\mu}$ in (\ref{gauge}), we say that
we have \emph{``fixed the gauge''}. In particular, an $A^\mu$ satisfying ( 
\ref{Lorenz}) is said \emph{`` to be in the Lorenz gauge''}. Still,
condition (\ref{Lorenz}) does not exhaust our liberty of choice, i.e., it
does not fix completely the $A^\mu$; we can go to the Lorenz gauge from any $%
A^\mu$ choosing a convenient $\phi$ such that it obeys 
\begin{equation}
\Box \phi+\partial_\mu A^\mu=0\Rightarrow \partial^{^{\prime}}_\mu A^\mu=0.
\end{equation}

A further transformation 
\begin{equation}
A^{^{\prime\prime}\mu}=A^{^{\prime}\mu}+\partial^\mu \phi^{^{\prime}},
\end{equation}
with $\phi^{^{\prime}}$ obeying 
\begin{equation}
\Box \phi^{^{\prime}}=0,
\end{equation}
will also lead us to $\partial_\mu A^{^{\prime\prime}\mu}=0$. So, a gauge
potential in the \emph{``Lorenz gauge''} will be determined except for a
gradient of an harmonic scalar field. This remnant or residual freedom can
be used to eliminate one of the components of $A^\mu$, such as, for example, 
$A^0$: Choose $\phi^{^{\prime}}$ such that 
\begin{equation}
\partial^0 \phi^{^{\prime}}=-A^{^{\prime}0},
\end{equation}
so that we have $A^{^{\prime\prime}0}=0$ for any space-time point $(t,\vec{x}
)$. Thus, $\partial_0 A^{^{\prime\prime}0}=0$ and the Lorenz condition will
then be 
\begin{equation}
\nabla \cdot \vec{A}=0\;; \qquad A^0=0.
\end{equation}

This gauge is known as the radiation gauge (or Coulomb one, $\nabla \cdot 
\vec{A}=0$). This gauge choice is not covariant, but can be realized in
every inertial reference frame.

This brings us to the analogy in the light-front case: 
\begin{eqnarray}
A^{^{\prime\prime}\mu}&=&A^{^{\prime}\mu}+\partial ^\mu \phi^{^{\prime}}; 
\notag \\
\partial^{+}\phi^{^{\prime}}&=&-A^{^{\prime}+}.
\end{eqnarray}

Therefore, $A^{^{\prime\prime}+}=A^{^{\prime}+}-A^{^{\prime}+}=0$, and we
obtain the following correspondence: 
\begin{eqnarray}
A^0=0 & \longrightarrow & A^+=0;  \notag \\
\nabla \cdot \vec{A}=0 & \longrightarrow & \partial^+ A^--\partial^\perp
A^\perp =0.
\end{eqnarray}

Note that the second equation above is the constraint (\ref{constraint}).
These imply the double Lagrange multipliers (terms for gauge fixing) in the
Lagrangian density herein proposed 
\begin{equation}
\mathcal{L}_{GF}=-\frac{1}{2\alpha}(n \cdot A)^2-\frac{1}{\beta}(\partial
\cdot A)^2.
\end{equation}

\end{document}